# On the Chances of Completing the Game of "Perpetual Motion"

Matthew C. Clarke[1] — July 2009

Email — matthew@clarke.name

For Kevin Prior — one of the world's Perpetual Motion Grand Masters

## Abstract

This brief paper describes the single-player card game called "Perpetual Motion" and reports on a computational analysis of the game's outcome. The analysis follows a Monte Carlo methodology based on a sample of 10,000 randomly generated games. The key result is that 54.55% ± 0.89% of games can be completed (by a patient player!) but that the remaining 45.45% result in non-terminating cycles. The lengths of these non-terminating cycles leave some outstanding questions.

## Introduction

Perpetual Motion [1] is a solitaire (i.e. single-player) game involving one standard pack of 52 playing cards. After shuffling the pack, a game proceeds as follows —

1. Four cards are laid side-by-side from left to right to form (or later, to add to) four piles.

2. If the four cards dealt all have the same value (suit is ignored throughout this game), then they are picked up and discarded.

3. If any two or three of the four cards dealt have the same value, then the duplicates are placed on top of the left-most card bearing the same value. (e.g. Suppose the cards dealt to the top of the four piles bear the values Ace, 3, Queen and 3 respectively. Then the top card of the fourth pile should be picked up and placed on top of the second pile.)

4. Steps 1 to 3 are repeated until all cards have been dealt.

5. Having gone through the whole pack, the four piles are now reformed into one pack by placing the first (left-most) pile on top of the second, the second on top of the third, and the third on top of the fourth.

6. Steps 1 to 5 (collectively called a *round*) are repeated *ad naseum*, until all cards are discarded in accordance with Step 2.

A player will soon notice two things. First, the game lasts for a long time, and is less frequently completed than terminated prematurely on account of boredom. Second, there is never any choice of moves and hence the outcome of the game is fully determined by the initial order of the shuffled cards. Nevertheless, there are languid Sunday afternoons when Perpetual Motion may seem slightly less dismal than the alternatives!

A third observation, which may take several games to discover, is that cycles sometimes develop. That is, once the number of cards has been significantly reduced, one finds that the order of card values at the end of a round is identical to the order at the end of some previous round. In such cases, the game can never be completed. (Note that, since we are ignoring the cards' suits, a cycle does not mean that the cards necessarily return to exactly the same order, only that the values on the cards return to the same order.)

The purpose of the current study is to examine the expected length of a game of Perpetual Motion. That is, how many rounds should one expect in a typical game? How many cards will have to be moved before the game is completed? In what percentage of games will a cycle appear?

---

# Methodology and Results

The complete population of shuffled packs consists of 52! elements, which is clearly too large to analyse exhaustively. However, following a Monte Carlo approach [2], the statistics gathered from a sizeable random sample of the shuffled packs can act as estimates for the whole population. A computer program which examined 10,000 games of Perpetual Motion yielded the following results.

Of the 10,000 games played, 5455 were completed and the remaining 4545 ended in a cycle. This is the combined result of ten disjoint samples of 1,000 games each. In order to establish the extent to which the overall result can be generalised to the whole population of possible games, we can consider the results from each of these ten samples to be normally distributed and examine the probability that the sample mean is the same as the population mean. With $\alpha=0.05$, we find that the population mean must lie within 8.96 of the sample mean. In other words, we can conclude that 54.55% ± 0.89% of games can be completed, while the other 45.45% will result in cycles.

Across the whole sample of 10,000 games, the average number of rounds required per game (whether played to completion or terminated as soon as a cycle was detected) was 128. The average number of rounds required before the first set of cards were discarded was 24. The average number of cards moved per game was 6201 (this includes dealing cards, moving them between piles and discarding them). Figure 1 shows the frequency distribution of the number of rounds in each game; Figure 2 shows the frequency distribution of the number of cards moved by the end of each game.

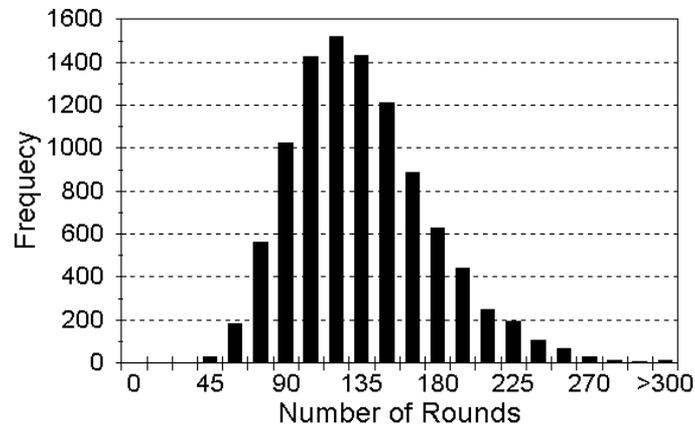

Figure 1. Frequency Distribution of Number of Rounds per Game

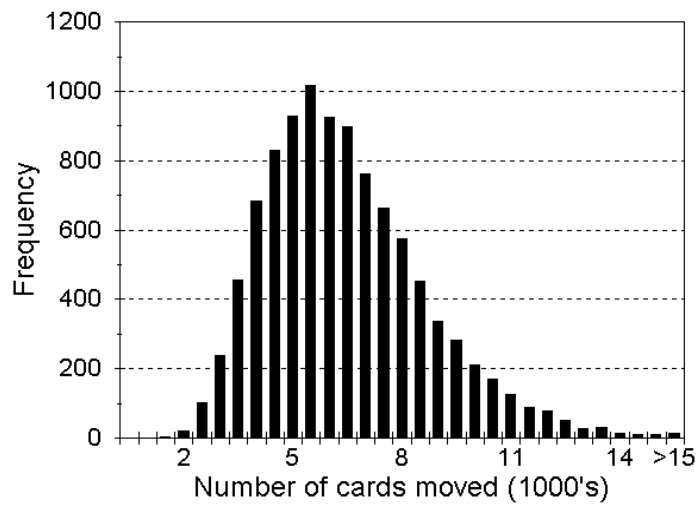

Figure 2. Frequency Distribution of Number of Cards Moved per Game

For those uncompletable games (i.e. those which resulted in a cycle), the frequency distribution of the cycle length is shown in Figure 3.

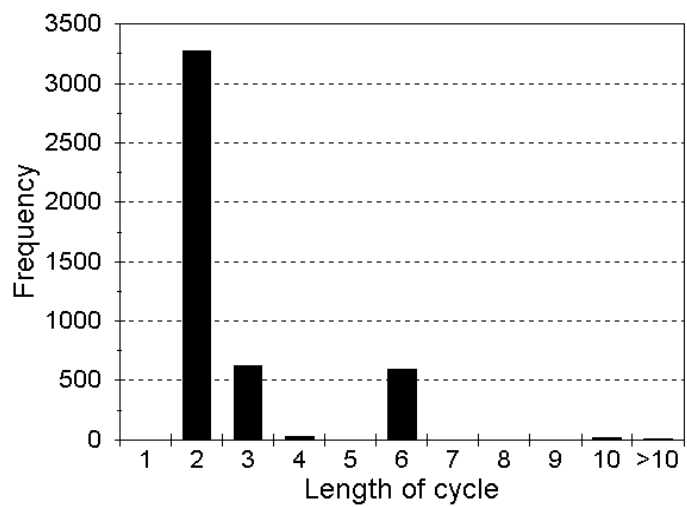

Figure 3. Frequency Distribution of Cycle Length in Uncompletable Games

## Discussion

The skewed distributions of the numbers of rounds and cards moved (Figures 1 and 2) are to be expected, but the distribution of cycle lengths is quite strange. The graph shows that by far the most common cycle is a 2-round cycle. When this occurs, it is normally when there are eight cards remaining, and it is easy to see how cycles of length 2 arise. There are also significant numbers of cycles of length 3 and (surprisingly) 6. Out of the 10,000 game sample, there were small numbers of cycles of length 4, 10, 12 and 24. Although no single-round cycles were found, there was one cycle of length 5 and one of length 8.

I cannot explain this pattern (or lack of pattern!), but I have hand-checking numerous examples and am convinced that the program's output is accurate. I cannot even explain why there are no single-round cycles. If playing a single round constituted a simple shuffle, then one would expect that the order after a round could never be the same as the order prior to that round. However, this is not applicable to this game for two reasons. First, since the card's suits are ignored, it may well be that a simple shuffle leaves the *values* of the cards in the same order, even if the actual cards have been rearranged. Second, playing one round of Perpetual Motion is not a simple shuffle since the final position of each card depends not just on the card's starting position, but also on the card's value. I would welcome any correspondence containing either an example of a single-round cycle, or a proof that no such cycles exist.

## Conclusion

A typical game of Perpetual Motion can be described as follows. After 24 rounds of play, the first set of four discards are removed. By the time the game is completed (approximately 55% of games) or a non-terminating cycle appears (45% of games), there will have been 128 rounds of play, requiring the player to move 6201 cards. If each move of a card takes one second, and recombining piles at the end of a round takes five seconds, then one should finish the game in slightly less than two hours of continuous play, just in time for dinner.

This, of course, assumes that the player will be sufficiently alert to notice when a cycle appears — an easy task for the vast majority of cases, but exceedingly difficult in games which lead to 24-round cycles!